\documentclass{PoS}

\usepackage{amsmath}
\usepackage{amssymb}
\usepackage{graphicx}

\makeatletter

\usepackage{fontenc}
\usepackage{times}
\usepackage{mathptmx}

\title{Monte Carlo studies on the expanding behavior 
of the early universe in the Lorentzian type IIB matrix model}

\ShortTitle{Monte Carlo studies on the early universe in the matrix model}

\author{\speaker{Yuta Ito}\\
Department of Particle and Nuclear Physics,\\
Graduate University for Advanced Studies (SOKENDAI),\\
Tsukuba, Ibaraki 305-0801, Japan\\
        E-mail: \email{yito@post.kek.jp}}

\author{Sang-Woo Kim\\
School of Physics, Korea Institute for Advanced Study (KIAS),\\
85 Hoegiro Dongdaemun-gu, Seoul 130-722, Korea\\
        E-mail: \email{sang@kias.re.kr}}

\author{Jun Nishimura\\
KEK Theory Center, High Energy Accelerator Research Organization,\\
Tsukuba, Ibaraki 305-0801, Japan\\
Department of Particle and Nuclear Physics,\\
Graduate University for Advanced Studies (SOKENDAI),\\
Tsukuba, Ibaraki 305-0801, Japan\\
        E-mail: \email{jnishi@post.kek.jp}}

\author{Asato Tsuchiya\\
Department of Physics, Shizuoka University,\\
836 Ohya, Suruga-ku, Shizuoka 422-8529, Japan\\
        E-mail: \email{satsuch@ipc.shizuoka.ac.jp}}

\abstract{
The type IIB matrix model is a conjectured nonperturbative 
formulation of superstring theory.
Recently the Lorentzian version of the model
has been studied by Monte Carlo simulation, and it has been shown that
only three out of nine spatial directions 
start to expand after a critical time.
We extend this work by investigating the expanding behavior 
for much longer time.
We find that the 3d space expands exponentially for some period of time, 
which may be interpreted as inflation.
We also simulate a simplified model, 
which is expected to capture some qualitative features of the original model
at much later times.
We observe that the exponential expansion eventually changes into a 
power-law ($t^{1/2}$) behavior, which agrees with
the expanding behavior of the 
Friedmann-Robertson-Walker (FRW) universe in the radiation dominated era.}

\FullConference{The 31th International Symposium on Lattice Field Theory,
                 Lattice2013\\
                 July 29 -- August 3, 2013\\
                 Mainz, Germany}
\makeatother

\begin{document}

\section{Introduction}

Understanding how our universe began is one of the most fundamental
themes in theoretical physics. 
For instance, it is widely believed that our
universe underwent a rapid expansion called inflation 
before the Big Bang. 
While there are many phenomenological models for inflation,
we have not yet understood it from
first-principle calculations in a fundamental theory.
Superstring theory is the most promising candidate
for such a fundamental theory, which can treat quantum gravity
and the Standard Model of particles in a unified manner.
Theoretical consistency requires that space-time should be 10d, 
but one can ``compactify'' the extra dimensions
to get our space-time without spoiling the consistency 
within perturbation theory.
The problem, however, is that there are actually too many consistent
backgrounds leading to different physics at low energy,
the situation which is commonly referred to as the Landscape nowadays.
On the other hand, if superstring theory can be formulated 
in a completely nonperturbative fashion,
as the lattice gauge theory does in QCD,
we may be able to obtain uniquely our 4d space-time 
with the Standard Model particles
propagating on it. 

The type IIB matrix model \cite{IKKT} is proposed 
as such a formulation in 1996.
An important feature of the model is that
the 10d space-time is described dynamically
as the eigenvalue distribution of the ten bosonic matrices 
$A_\mu$ ($\mu=0,1,\cdots ,9$). 
In particular, by identifying the dominant matrix configurations
in the partition function, one can investigate
what kind of space-time is generated dynamically in this model.
Until quite recently, this issue has been addressed
in the Euclidean version of the model, in which the temporal matrix
is Wick-rotated as $A_0 = - i A_{10}$.
The space-time represented in such a model is then actually Euclidean
and, in particular, one cannot apply it to cosmology
since one cannot extract the real-time dynamics. 

In 2011 three of the authors (S.-W.K, J.N.\ and A.T.)
studied the Lorentzian version of the type IIB matrix model 
for the first time by Monte Carlo simulation \cite{KNT1}.
The real-time evolution was extracted
from the dominant matrix configurations, and it was found
that 3 out of 9 spatial directions start to expand 
after a critical time. 
There are also other recent developments in the type IIB matrix model.
Refs.~\cite{Aoki,CSZ,NT1,NT2} discussed how 
to realize the Standard Model in the type IIB matrix model
and extended models, while
refs.~\cite{Steinacker,KNT2,Kim:2012mw} discussed classical solutions 
in the Lorentzian model, which are consistent with our 4d space-time.

In this paper we extend the Monte Carlo studies in ref.~\cite{KNT1} 
by studying the expanding behavior for much longer time. 
First we find that the expansion in three directions is
actually exponential, which may be interpreted as the beginning
of inflation. This behavior is confirmed 
with larger matrix size
in a simplified model, which can be obtained by 
keeping only
the term proportional to the temporal matrix $A_0$
in the fermionic action.
This simplification emphasizes the effects of fermionic matrices
which cause a repulsive force between the eigenvalues
of the temporal matrix $A_0$.
Such effects are expected to become less important as the universe
expands due to
the term proportional to the spatial matrices $A_i$ ($i=1,\cdots ,9$)
in the fermionic action.
Therefore, as a simplified model which is expected to capture 
qualitative behaviors at late times,
we study the quenched model, which is obtained by
simply omitting the fermionic matrices.
We find in this model that the expansion is exponential
for some time, but then it changes into a power law ($t^{1/2}$),
which agrees with the expanding behavior 
of the FRW universe in the radiation dominated era.

The rest of this paper is organized as follows. 
In section \ref{sec:definition} 
we define the Lorentzian version of the type IIB matrix model. 
In section \ref{sec:early-time} we show that 
the expansion in three directions turns out to be exponential.
This behavior is also reproduced 
with larger matrices
by a simplified model for early time behaviors.
In section \ref{sec:late-time} we study yet another simplified
model, which is expected to capture 
qualitative features
at late times, and show that the exponential expansion 
changes into a power-law behavior at some point in time. 
Section \ref{sec:summary} is devoted to a summary and discussions.

\section{Lorentzian version of the type IIB matrix model}
\label{sec:definition}

The type IIB matrix model \cite{IKKT} is defined 
in its Lorentzian version
by the partition function \cite{KNT1}
\begin{eqnarray}
Z & = & \int dA \, d\Psi\, e^{i\left(S_{{\rm b}}+S_{{\rm f}}\right)}
\ ,
\label{eq:Pf}
\end{eqnarray}
where the action is given by 
\begin{eqnarray}
\label{Sb-def}
S_{{\rm b}} & = & 
-\frac{1}{4g^{2}} \, {\rm Tr}
\left(\left[A_{\mu},A_{\nu}\right]\left[A^{\mu},A^{\nu}\right]\right)
 \ ,\\
S_{{\rm f}} & = & 
-\frac{1}{2g^{2}} \, {\rm Tr}\left(\Psi_{\alpha}
\left(C\Gamma^{\mu}\right)_{\alpha\beta}
\left[A_{\mu},\Psi_{\beta}\right]\right) \ .
\label{Sf-def}
\end{eqnarray}
We have introduced $N\times N$ traceless Hermitian matrices
$A_{\mu}\left(\mu=0,\cdots,9\right)$ 
and $\Psi_{\alpha}\left(\alpha=1,\cdots,16\right)$, which
are bosonic and fermionic, respectively.
The Lorentz indices $\mu$ and $\nu$ are contracted 
using the metric $\eta={\rm diag}\left(-1,1,1,\cdots\right)$.
$\Gamma^{\mu}$ are 10d gamma matrices after the Weyl projection
and $C$ is the charge conjugation matrix. 
The parameter $g$ in (\ref{Sb-def}) and (\ref{Sf-def})
can be absorbed by rescaling $A_{\mu}$ and $\Psi_{\alpha}$. 
The model has SO(9,1) Lorentz symmetry as well as SU($N$) symmetry. 

One finds that the bosonic action is proportional to
\begin{equation}
S_{{\rm b}}  \propto  {\rm Tr}\left(F_{\mu\nu}F^{\mu\nu}\right)
=
-2 \, {\rm Tr}
\left(F_{0i}\right)^{2}+{\rm Tr}\left(F_{ij}\right)^{2} \ ,
\label{Sb-decompose}
\end{equation}
where we have defined Hermitian matrices 
$F_{\mu\nu}=i\left[A_{\mu},A_{\nu}\right]$.
Therefore,
the bosonic action is not positive definite.
In order to make the partition function finite,
one actually needs to introduce infrared cutoffs
\begin{eqnarray}
\frac{1}{N}\, {\rm Tr}\left(A_{0}\right)^{2} & \leq & 
\kappa L^{2},\label{eq:cutoff t}\\
\frac{1}{N} \, {\rm Tr}\left(A_{i}\right)^{2} & \leq & 
L^{2} 
\label{eq:cutoff s}
\end{eqnarray}
in both temporal and spatial directions.
It turned out that 
these cut-offs can be removed in the large-$N$ limit,
and clear scaling behaviors corresponding to the continuum and 
infinite-volume limits were observed \cite{KNT1}.
This implies that the resulting theory has no parameters
except the scale parameter.
In actual simulation, we set $L=1$ without loss of generality 
since it only fixes the scale,
and choose $\kappa$ appropriately as a function of $N$
so that both the continuum and infinite-volume limits are taken.

The partition function (\ref{eq:Pf}) is not suitable
for Monte Carlo simulation due to the phase factor
$e^{i S_{{\rm b}}}$.
However, 
by integrating out the scale factor of the bosonic matrices,
one can rewrite the partition function into the form
that allows direct Monte Carlo studies 
without the sign problem \cite{KNT1}
\begin{equation}
Z=\int dA\,{\rm Pf}\mathcal{M}\left(A\right)\delta
\left(\frac{1}{N}{\rm Tr}\left(F_{\mu\nu}F^{\mu\nu}\right)\right)
\delta\left(\frac{1}{N}{\rm Tr}\left(A_{i}\right)^{2}-L^{2}\right)
\theta\left(\kappa L^{2}-\frac{1}{N}{\rm Tr}\left(A_{0}\right)^{2}\right)\ ,
\label{eq:IIBMM p.f.}
\end{equation}
where $\theta\left(x\right)$ represents the step function.
The Pfaffian ${\rm Pf}\mathcal{M} (A)$ 
in (\ref{eq:IIBMM p.f.}),
which is obtained by integrating out fermionic matrices, 
is real in the present Lorentzian case, and it does not cause 
any sign problem.

\begin{figure}[t]
\centering{}
\includegraphics{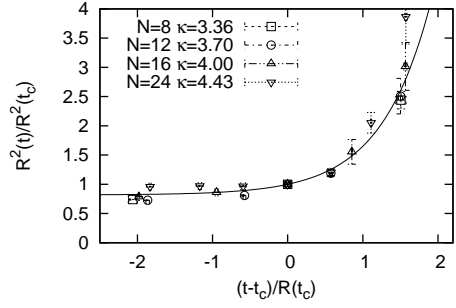}
\caption{\label{fig:1}The extent of space 
$R^2\left(t\right)/R^2\left(t_{c}\right)$
is plotted against $\left(t-t_{c}\right)/R\left(t_{c}\right)$ 
for the original model 
(\protect\ref{eq:IIBMM p.f.})
with various $\kappa$ and $N$.
%
The solid line is a fit to the exponential behavior
$y=a+ (1-a) \exp (bx)$ with $a=0.82(1)$ and $b=1.5(2)$.
}
\end{figure}

In order to extract the time evolution from configurations
generated by (\ref{eq:IIBMM p.f.}),
we first diagonalize the temporal matrix $A_{0}$ as
\begin{equation}
A_{0}={\rm diag}\left(\alpha_{1},\cdots,\alpha_{N}\right) \ ,
\quad\quad {\rm where~} \alpha_{1}<\cdots<\alpha_{N} \ ,
\end{equation}
using the SU$\left(N\right)$ symmetry.
In such a basis, it turned out that the spatial matrices $A_{i}$
have a band-diagonal structure; namely it was found that
the off-diagonal elements $\left(A_{i}\right)_{IJ}$
with $\left|I-J\right|>n$ are small for some $n$.
This non-trivial dynamical property motivates us to
define $n\times n$ matrices\footnote{The value of $n$ should be chosen
appropriately by measuring the fall-off of the off-diagonal elements.
In fig.~\ref{fig:1} we use $n=N/4$;
in fig.~\ref{fig:2} we use $n=8$;
in fig.~\ref{fig:3} we use $n=8$ except for $N=128$, where we use $n=12$.}

\begin{equation}
\left(\bar{A}_{i}(t)\right)_{ab}
\equiv\left(A_{i}\right)_{\nu+a,\nu+b} \ ,
\end{equation}
where $\nu=0,1,\cdots,N-n$, and $a,b=1,\cdots , n$. 
We consider that these block matrices
represent the states of the universe at time $t$, where
\begin{equation}
t=\frac{1}{n}\sum_{a=1}^{n}\alpha_{\nu+a} \ .
\end{equation}
The basic quantity we calculate in this paper is
the extent of space at time $t$ defined as
\begin{equation}
R^{2}(t)= \frac{1}{n}\, {\rm tr}\left(\bar{A}_{i}(t) \right)^{2} \ .
\label{eq:R(t)}
\end{equation}

\section{Exponential expansion at early times}
\label{sec:early-time}

First we study the model \eqref{eq:IIBMM p.f.}
by Monte Carlo simulation. 
In fig.\ref{fig:1}
we plot the extent of space (\ref{eq:R(t)}) as a 
function of time $t$.
Here and hence forth, we normalize dimensionful quantities by
$R(t_{\rm c})$, where $t_{\rm c}$ is the ``critical time''
at which the spatial SO(9) symmetry is spontaneously
broken down to SO(3) and only three out of nine spatial directions
start to expand \cite{KNT1}.
Compared with the previous work \cite{KNT1}, we were able to simulate
larger matrices and hence a longer time period.
In fact the obtained $R(t)$ can be nicely fitted 
with $y=f(x)\equiv a+ (1-a) \exp (bx)$,
where we have imposed $f(0)=1$, 
which follows from the chosen normalization.
This implies that three spatial directions actually start to
expand exponentially, which may be interpreted as the beginning
of inflation.

\begin{figure}
\centering{}
\includegraphics{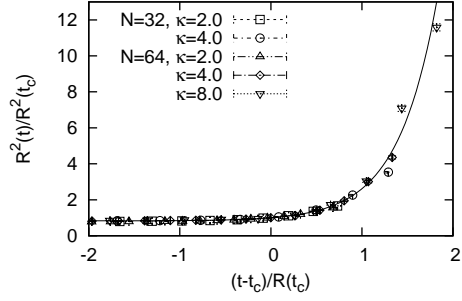}
\caption{\label{fig:2}The extent of space 
$R^{2}\left(t\right)/R^{2}\left(t_{\rm c}\right)$
is plotted against 
$\left(t-t_{\rm c}\right)/R\left(t_{\rm c}\right)$ 
for the simplified model for early times 
in the $d=5$ case with various $\kappa$ and $N$.
The solid line is a fit to the exponential behavior
$y=a+ (1-a) \exp (bx)$ with $a=0.83(1)$ and $b=2.3(1)$.
}
\end{figure}


In order to confirm the exponential behavior for 
a longer time period,
we need to increase the matrix size further, which makes 
the simulation too time-consuming.
Here we consider, instead, a simplified model that describes the behavior
at early times.
For that we decompose the fermionic action (\ref{Sf-def}) into
two terms as
\begin{equation}
S_{{\rm f}} 
 \propto 
{\rm Tr}\left(\Psi_{\alpha}
\left(C\Gamma^{0}\right)_{\alpha\beta}
\left[A_{0},\Psi_{\beta}\right]\right)+
{\rm Tr}\left(\Psi_{\alpha}\left(C\Gamma^{i}\right)_{\alpha\beta}
\left[A_{i},\Psi_{\beta}\right]\right) \ .
\label{eq:fermionic action}
\end{equation}
Due to the expanding behavior of the universe,
the elements of the spatial matrices $A_i$ become very large
at late times.
At early times, on the other hand,
it is expected that the first term 
in (\ref{eq:fermionic action}) is more important,
so we simply omit the second term 
in \eqref{eq:fermionic action} as a simplification. 
Integrating out the fermionic matrices\footnote{Strictly speaking,
there are zero modes corresponding to $\Psi_\alpha$
satisfying $[A_0 , \Psi_\alpha] = 0$, which we simply neglect.}, 
we obtain the Pfaffian, which is now given by 
\begin{equation}
{\rm Pf}\mathcal{M}\left(A\right)=\Delta^{2(d-1)} \ , 
\label{eq:Pf early}
\end{equation}
where $\Delta \equiv \prod_{i>j}\left(\alpha_{i}-\alpha_{j}\right)$ 
is the van der Monde determinant
and we have written down the general results for dimensionally
reduced SYM models with $d$ spatial dimensions 
($d=9$ in the case of type IIB matrix model). 
The Pfaffian \eqref{eq:Pf early} obtained here causes
a repulsive force between all the pairs of eigenvalues of $A_{0}$,
which cancels the attractive force arising from
the fluctuation of the bosonic matrices at the one-loop level.
Due to this cancellation,
the eigenvalues of $A_{0}$ can extend to infinity, which 
necessitates the cutoff \eqref{eq:cutoff t} in the temporal direction. 

The simplified model for early times
with the Pfaffian replaced by (\ref{eq:Pf early})
can be simulated with much less efforts.
Here we study the $d=5$ model,
in which the rotational SO(5) symmetry is broken down 
to SO(3) at some critical time $t_{\rm c}$
analogously to the $d=9$ model.
In fig.~\ref{fig:2} we plot the extent of space
\eqref{eq:R(t)}
as a function of $t$ for various $N$ and $\kappa$. 
This confirms the exponentially expanding behavior
in the simplified model,
which suggests that 
the first term of 
the fermionic action \eqref{eq:fermionic action}
is indeed important for the space to expand exponentially.

\section{Power-law expansion at late times}
\label{sec:late-time}

At late times, 
the second term in the fermionic
action (\ref{eq:fermionic action}) 
becomes more important, and it is expected that 
the repulsive force represented by (\ref{eq:Pf early})
is no more effective. In order to mimic such a situation,
we consider a quenched model obtained by omitting the
fermionic matrices completely.
In this model,
since the eigenvalues of $A_{0}$ attract each other,
we do not introduce the cutoff (\ref{eq:cutoff t}) 
in the temporal direction.
The extent of the eigenvalue distribution 
increases with $N$, however, 
and one can take both the continuum and infinite-volume limits.
The breaking of SO(5) symmetry (for $d=5$)
down to SO(3) is observed 
after a critical time $t_{\rm c}$
for sufficiently large matrix size $N$.

\begin{figure}
\centering{}
\includegraphics{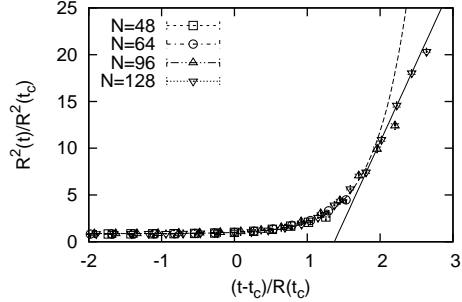}
\caption{\label{fig:3}The extent of space 
$R^{2}\left(t\right)/R^{2}\left(t_{\rm c}\right)$
is plotted against 
$\left(t-t_{\rm c}\right)/R\left(t_{\rm c}\right)$ 
for the $d=5$ quenched model with various $N$,
which is considered as a simplified model for late times.
The dashed line represents a fit
$y=a+\left(1-a\right)e^{bx}$
to the early time behavior ($a=0.870(3)$, $2.21(3)$), 
whereas the solid line represents
a fit $y=cx+d$ to the late time behavior ($c=17.0(1)$, $d=-23.3(3)$).}
\end{figure}

In fig.~\ref{fig:3} we plot the extent of space
\eqref{eq:R(t)} for the $d=5$ quenched model.
The exponential behavior is observed for some period after the
critical time, but it changes into a linear behavior
$R^{2}(t) \sim t$ meaning that $R (t) \sim t^{1/2}$,
which agrees with the expanding behavior
of the 
FRW
universe in the radiation dominated era.

\section{Summary and discussions}
\label{sec:summary}

In this paper we investigated the expanding behavior of 
the early universe in the Lorentzian version of the type
IIB matrix model, which is considered to be a nonperturbative
formulation of superstring theory.
First we studied the original model
for a longer time period
than in the previous study \cite{KNT1},
and found that three out of nine spatial directions actually 
start to expand \emph{exponentially} after a critical time,
which may be interpreted as the inflation. 

In order to study the behavior for much longer time,
we considered two simplified models for early times and 
for late times, 
respectively.
The model for early times is defined
by omitting the term proportional to $A_{i}$ in the fermionic action. 
This simplification emphasizes
the repulsive force between the eigenvalues of $A_{0}$,
which cancels the attractive force due to the fluctuation of the 
bosonic matrices.
Indeed 
the exponential expansion was confirmed
in this model suggesting the important role played 
by the repulsive force
due to the fermionic matrices at early times.
This motivated us to define
a simplified model for late times by the quenched model,
in which the repulsive force is absent.
It turned out that the eigenvalue distribution of $A_{0}$
is finite without the cutoff (\ref{eq:cutoff t}),
but its extent increases with $N$.
In particular, the SSB from SO(5) to SO(3) occurs after a critical time
for sufficiently large $N$.
We find that the expansion behavior changes
from an exponential one to a power law $t^{1/2}$ at some time
after the critical time. This power-law expansion agrees with
that of the radiation dominated FRW universe.

As future prospects, it would be interesting to study the quenched model
with larger matrices to see whether the power law changes, for instance,
into the one for the matter dominated era at later times.
Along the same line, we can see whether the
universe somehow cools down and whether
the classical equations of motion become valid at late times
as conjectured in ref.~\cite{KNT2,Kim:2012mw}.
Finally, it would be important to confirm 
the transition from the exponential behavior
to the power-law behavior directly in the original model.
In particular, this will give us the value of E-folding,
which is determined \emph{dynamically}
in the Lorentzian type IIB matrix model.

\section*{Acknowledgment}

Computation was carried out on PC clusters at KEK
and supercomputers SR16000 at YITP, Kyoto University 
and FX10 at University of Tokyo.
The work of 
Y.~I.\ is supported by Grant-in-Aid for 
JSPS
fellows.
The work of S.~-W.~K.\
is supported by the National Research Foundation
of Korea (NRF) Grant funded by the Korean Government
(MEST 2005-0049409 and NRF-2009-352-C00015).
The work of J.~N.\ and A.~T.\ is supported
by Grant-in-Aid for Scientific
Research
(No.\ 20540286, 24540264, and 23244057)
from JSPS.

\end{document}